# Achieving Optical Refractive Index of 10-Plus by Colloidal Self-Assembly

NaYeoun Kim[1,+], Ji-Hyeok Huh[1,2,+], YongDeok Cho[1,+], Sung Hun Park[1], Hyeon Ho Kim[1], Kyung Hun Rho[1], Jaewon Lee[1], and Seungwoo Lee[1,3,4*]

[1]KU-KIST Graduate School of Converging Science and Technology, Korea University, Seoul 02841, Republic of Korea

[2]Department of Applied Physics, Hanyang University, Ansan 15588, Republic of Korea

[3]Department of Integrated Energy Engineering (College of Engineering), Department of Biomicrosystem Technology, and KU Photonics Center, Korea University, Seoul 02841, Republic of Korea

[4]Center for Opto-Electronic Materials and Devices, Post-Silicon Semiconductor Institute, Korea Institute of Science and Technology (KIST), Seoul 02792, Republic of Korea

*Email: seungwoo@korea.ac.kr

**Abstract**: This study demonstrates the developments of self-assembled optical metasurfaces to overcome inherent limitations in polarization density ($P$) within natural materials, which hinder achieving high refractive indices ($n$) at optical frequencies. The Maxwellian macroscopic description establishes a link between $P$ and $n$, revealing a static limit in natural materials, restricting $n$ to approximately 4.0 at optical frequencies. Optical metasurfaces, utilizing metallic colloids on a deep-subwavelength scale, offer a solution by unnaturally enhancing $n$ through electric dipolar (ED) resonances. Self-assembly enables the creation of nanometer-scale metallic gaps between metallic nanoparticles (NPs), paving the way for achieving exceptionally high $n$ at optical frequencies. This study focuses on assembling polyhedral gold (Au) NPs into a closely packed monolayer by rationally designing the polymeric ligand to balance attractive and repulsive forces, in that polymeric brush-mediated self-assembly of the close-packed Au NP monolayer is robustly achieved over a large-area. The resulting monolayer of Au nanospheres (NSs), nanooctahedras (NOs), and nanocubes (NCs) exhibits high macroscopic integrity and crystallinity, sufficiently enough for pushing $n$ to record-high regimes. The study underlies the significance of capacitive coupling in achieving an unnaturally high $n$ and explores fine-tuning Au NC size to optimize this coupling. The achieved $n$ of 10.12 at optical frequencies stands as a benchmark, highlighting the potential of polyhedral Au NPs in advancing optical metasurfaces.

## 1. Introduction

The Maxwellian macroscopic description implies that the polarization density ($P$) of the constituent building blocks in the natural matter, such as atoms and molecules, plays a crucial role in determining the refractive index ($n = \sqrt{1 + (P/\varepsilon_0 E)}$, where $\varepsilon_0$ and $E$ correspond to the effective relative permittivity and electric field, respectively)(*1–4*). Due to the size mismatch between atoms/molecules and the wavelength of interest, the achievable $P$ from naturally occurring materials is constrained by a static limit at optical frequencies (spanning from the visible to near-infrared (NIR)). This statically limited $P$ results in a relatively low upper limit of $n$ (~ 4.0)(*5, 6*). The fundamental limitations of $P$ in this natural regime can be overcome through the concept of optical metasurfaces. The deep-subwavelength scale and high optical conductivity of synthetically produced or lithographically defined metallic nanoparticles (NPs with the size of < 100 nm) enable an unnatural enhancement of $n$ in the host medium (i.e., a metallic NPs-dispersed homogeneous medium) via harmonically induced electric dipolar resonances. In this context, the metallic NPs can be regarded as 'artificial optical atom equivalents (i.e., meta-atoms)'(*7–12*).

More importantly, the capacitive coupling between resonantly polarized electric dipoles (EDs) can further elevate the available $P$ and $n$ into an extraordinarily high regime. Once the size and element of such metallic NPs are determined, the distance between each NP becomes a critical factor governing the accessible range of capacitance and the resulting $P$. In line with this, chemically synthesized colloidal NPs offer advantages over alternative methodological counterparts (e.g., lithographically defined NPs), because the soft self-assembly of colloids into a closely packed monolayer facilitates the creation of a few nanometer-scale metallic gaps between NPs across a large area, which is otherwise unattainable(*13–15*). Overall, colloidal self-assembly has paved the way for achieving an unnaturally high-$n$ at optical frequency(*5, 6, 11, 12, 16–22*).

It is also noteworthy that colloidal polyhedra, serving as soft meta-atoms, offer advantages over spheres. This is attributed to the fact that, within nanogaps, the strength of capacitive coupling can be enhanced by the facet, as opposed to a point-like interfacial morphology (*7, 15, 17, 23*). Recent advancements in chemical synthetic routes have translated the concept of highly uniform metallic polyhedra from imagination into reality(*24–26*). However, the 2D self-assembly of metallic polyhedra, particularly in the context of achieving high-$n$ optical metasurfaces, has generally resulted in low macroscopic lattice integrity and crystallinity. This limitation, in turn, has restricted the upper limit of achievable $n$, approximately 6.4(*17*).

In this study, we use polymeric brush-assisted balancing of the attractive and repulsive potentials of gold (Au) polyhedra to assemble a large-scale, closely packed monolayer of Au polyhedra with high macroscopic integrity and crystallinity. These quantities are crucial for achieving an extremely high $n$. In particular, the polymeric ligands used (i.e., thiol-terminated polystyrene, abbreviated as PSSH) are rationally designed to coat the surfaces of Au colloids, encoding repulsive forces that slightly overwhelm the intrinsically strong van der Waals ($W_v$) attractions of Au NPs. Regardless of the shapes and sizes of the Au colloids, the 2D interfacial confinement of these polymer-grafted Au NPs leads to their self-assembly into the 2D monolayer. The Au NPs used in this study include Au nanospheres (Au NSs), Au nanooctahedras (Au NOs), and Au nanocubes (AuNCs), with their size uniformity maintained from 50 to 60 nm. As such, we can experimentally constrain the determinant of $n$ to one degree of freedom, i.e., the interfacial morphology between Au NPs. We systematically enhance the maximally (resonantly) achievable $n$ of the Au NP 2D superlattice in the order of Au NSs, NO, and NCs. Critically, our findings reveal that a closely packed monolayer of Au NCs can push $n$ into record-high regimes (10.12 for $n$), which were previously out of reach at optical frequencies.

## 2. Results and Discussion

### 2.1 Synthesis and Assembly

As presented in **Fig. 1A-C**, a collective set of Au colloids used in this study, ranging from 50 to 60 nm in size and consisting of highly uniform Au NSs, NO, and NCs, were synthesized using the 'iterative etching and growth' method, as depicted in **fig. S1** (for more details, see **Supplementary Information**). A statistical analysis of the size distribution validated the uniformity of the synthesized Au NPs (**fig. S2**). The as-synthesized Au NPs were coated with cetyltrimethyl ammonium bromide (CTAB). Their Zeta potential ranged from -10 mV to -20 mV: for example, -18.74 mV and -11.28 mV for Au NSs and NCs, respectively (see the left panel of **Fig. 1D**). Toward the 2D confined assembly of these Au colloids(*14, 27, 28*), two different methods were undertaken for the 2D confined Au colloidal self-assembly, as depicted in **Fig. 1E-F**. The Au colloidal aqueous solution was initially top-coated onto the hexane oil fluids. The subsequent introduction of ethanol into the aqueous solution spurred the Au NPs to be confined at the interface between hexane oil and water layers (see **Fig. 1E**). This interfacial assembly was promoted by polymeric brush-boosted entropy along with Van der Waals-mediated enthalpy. In particular, the Zeta potential of CTAB-coated Au NPs decreased to near zero (~ -0.06 mV for Au NSs; ~ -0.28 mV for Au NCs) due to the addition of ethanol (see the left panel of **Fig. 1D**). This led to excessively strengthened $W_v$, resulting in a kinetically trapped, diffusion-limited fractal growth, as shown in a high-magnification transmission electron microscope (TEM) image (**Fig. 1G**). The resulting 2D Au NC monolayer showed poor integrity and crystallinity.

Instead of a single molecular ligand, a polymer brush was utilized to improve the Zeta potential further. To achieve this, we used PSSH with a minimum molecular weight ($M_w$) of 5K because the shorter PSSH chains, for example, a molecular weight of 2K, could not maintain a stable colloidal suspension due to insufficient polymeric repulsion force (see **fig. S4**). Following the ligand exchange from CTAB to PSSH, the Zeta potentials of Au NSs and NCs increased to -39.85 mV and -27.60 mV, respectively (see the left panel of **Fig. 1D**). The UV-Vis absorption spectra evidenced this ligand exchange (**fig. S3**). The PSSH-coated Au NPs, dispersed in toluene, were coated onto a diethylene glycol (DEG) subphase via the drop-casting method. Then, complete evaporation of toluene spontaneously pushed the PSSH-coated Au NPs into a closely packed 2D monolayer. In this modified assembly, the presence of a polymer brush physically repels the Au NPs, and their repulsive potentials ($W_r$) can overwhelm $W_v$. The corresponding potential diagram, quantitated by Derjaguin-Landau-Verwey-Overbeek (DLVO) theory, is presented in the right panel of **Fig. 1D**. During this assembly, the net potential energy of repulsive interaction is significantly higher than the attractive interaction regardless of the distance between Au NPs. Therefore, the reaction-limited 2D assembly of Au NPs becomes feasible, resulting in significantly improved macroscopic integrity and crystallinity (**Fig. 1H**).

It is noteworthy that Au NCs showed a weaker Zeta potential as compared to Au NSs. This discrepancy originated from the sequestration of organic ligand attachments on a faceted surface, which is more pronounced than on a spherical surface(*26*). The effect of steric hindrance becomes more dominant with the increased flatness of the target surface. Nevertheless, the experimentally obtained $W_v$ across all the Au NPs proved to be adequate for the self-assembly of 2D monolayers of

Au NSs, NOs, and NCs with high macroscopic integrity, as shown in scanning electron microscope (SEM) images (**Fig. 2A-C**).

In particular, Au NSs (**Fig. 2A**) and Au NOs (**Fig. 2B**) presented a 6-fold hexagonal packing arrangement, while Au NCs (**Fig. 2C**) showed a 4-fold square lattice packing arrangement (see more details in **fig. S5**). The inset images within each SEM elucidate the orientation angles, $\theta_n$, between neighboring NPs, where subscript $n$ corresponds to the number of nearest neighbors (also see **fig. S6**). Given the angle $\theta_n$ quantified from such SEM insets, we mapped Voronoi diagrams of all the assembled Au NP monolayers, where the monolayers' degree of order, $\Psi_n$, is spatially quantified with a different false color, ranging from 0 to 1 (for more details, see **Supplementary Information**)(*29, 30*). The higher value of $\Psi_n$ implies the higher degree of macroscopic integrity. For both the hexagonally packed Au NSs and Au NO monolayers, $\Psi_6$ was calculated, yielding values of 0.72 and 0.57, respectively (**Fig. 2D-E**). For the square lattice-arranged Au NCs, $\Psi_4$ exhibited a value of 0.41 (**Fig. 2F**). The macroscopic crystal integrities of the assembled Au NP monolayers were found to be increased in the order of Au NCs, NOs, and NSs, because higher coordination number can reduce more enthalpy during close-packing of NPs.

The numerical simulation results, supported by HOOMD(*31, 32*), further confirm our experimental observations, providing a theoretical framework for the self-assembly patterns of Au NSs, Au NO, and Au NCs (**Fig. 2G-I**). Especially, the Voronoi diagrams derived from the HOOMD simulation data (**Fig. 2J-L**) show good agreements with the experimental findings, with the $\Psi_n$ values for assemblies' order being 0.85, 0.59, and 0.40 for AuNSs, AuNO, and AuNCs respectively. These simulation results support the reliability of the experimentally observed packing rule in **Fig. 2A-C** (for more details, see **Supplementary Information**). Even if Au NCs showed the lowest integrity of crystals, the few tens of micron-scale dimensions for each domain were still obtained reliably. As detailed below, these crystal domain sizes are sufficient for utilizing Au NC monolayer as an optically effective medium, allowing us to macroscopically and reliably characterize effective optical parameters (i.e., $P$ and $n$), and achieving unnaturally high $n$ of 10-plus.

## 2.2 Colloidal shape dependency on refractive index

The Lorentzian resonant oscillatory behaviors of the Au NP monolayers can be confirmed by measuring bright field, normal reflection spectra. In the case of a polyhedral monolayer, the ED resonances can be evidenced by the presence of a dark-mode, resulting from the destructive coupling between two oppositely induced EDs at the top and bottom of the Au NP monolayer(*22, 33, 34*). This dark mode gives rise to the spectral dip in normal reflection spectra. Modal analyses of Au NO and NC monolayers at ED modes, supported by finite-element method (FEM) numerical calculations, highlight these main features of the dark-mode, whereas a single ED can harmonically oscillate and radiate (bright mode) in an Au NS monolayer (see **Fig. 3A-C**; also see the Supplementary Movies). Moreover, the spectral position of this normal reflection dip enables the quantification of the nanogap between Au NPs in a closely packed monolayer and cross-verifies the nanogap quantitated from the TEM images in **Fig. 1H**, and vice versa(*17*). For example, a smaller gap gives rise to a stronger capacitive coupling between EDs, as evidenced by a redshifted spectral dip in normal reflection.

**Fig. 3D-F** show the measured normal reflection spectra across Au NS (**Fig. 3D**), NO (**Fig. 3E**), and NC (**Fig. 3F**) monolayers (solid lines), well coinciding with theoretical predictions (dotted lines). The nanogaps quantified from these normal reflection spectra for all Au NP monolayers consistently measured around 6 nm, matching with the results obtained from TEM analyses. These reflection spectra further validate the high macroscopic integrity and crystallinity of the assembled Au NP monolayers.

Given the defined nanogap, we then experimentally characterized and systematically compared the $n$-dispersions across the Au NP monolayers, as shown in **Fig. 3G-I** (dotted and solid lines, respectively, for experimental and modified Drude-Lorentz fitting curve). Herein, the $n$-dispersions were measured with the assistance of general ellipsometry (see more details in **Supplementary Information**). As expected, the achieved $n$ values at resonant wavelengths were increased in the order of Au NSs ($n$ of 5.3), NOs ($n$ of 6.2), and NCs ($n$ of 7.0), surpassing what naturally occurring materials can exhibit (~ 4.0). At off-resonant wavelengths (i.e., 1600 nm wavelengths), the values of 3.3, 3.6, and 3.9 were comparable to or slightly lower than the natural upper limit. As with normal reflection spectra, both experimental results and theoretical predictions were in good coincidence. Consequently, given these obtained $n$-dispersions, we can numerically extract the dimensionless Lorentz parameter $s$ (generally referred to as the effective oscillator strength) to quantify the achievable strength of ED resonance (for more details, see eq. 5 of **Supplementary Information**). In contrast to $P$, $s$ is related to the effective oscillator strength of the meta-atoms, independent of the working frequency. Thus, $s$ allows us to systematically compare the ED resonance strength regardless of the resonant frequency. As expected, Au NCs outperform Au NSs and NOs in terms of pushing the upper limit of $s$ to the higher regimes (i.e., $s$ of 4.76 for Au NSs, $s$ of 6.02 for Au NOs, and $s$ of 8.74 for Au NCs), because given the size of Au NPs (e.g., 50 nm), the interfacial area of NCs can be larger than those of NSs and NOs.

### 2.3 Reaching benchmark optical refractive index (10-plus)

We then fine-tuned the size of the Au NC to further enhance the strength of the resonantly induced EDs and their capacitive coupling, because a larger Au NC can redshift its ED resonance wavelength, further moving away from the interband transition regime. However, there is a technical conundrum, since a thickened Au NC monolayer can also improve a diamagnetic effect, compromising the possible upper limit of the $n$(*3, 17*). In our experiments, 60 nm-sized Au NCs allowed us to achieve the maximum $n$ at the ED resonant regime, as summarized in **Fig. 4A-D**. In particular, a selective increase in $M_w$ of PSSH (from 5K to 50K), grafted on the same-sized 60 nm Au NC, deterministically increased the nanogap (from 5 nm to 10 nm), while the interfacial area between Au NCs remained almost intact. As expected, a longer PSSH (resulting in a larger nanogap) led to weakened capacitive coupling, as highlighted by a blue-shifted ED resonance wavelength, the lower $s$ (**Table S2**, **Supplementary Information**), and a reduced $n$ at the resonant regime. Notably, as shown in **Fig. 4D**, the monolayer composed of 60 nm Au NCs with 5K PSSH showed the highest $n$ of 10-plus. It is noteworthy that the achieved $s$ of 29.06 and $n$ of 10.12 are the highest values at optical frequencies among others (**Fig. 4E**). Lastly, note that this polymeric brushed mediated interfacial assembly allows the robust and reliable development of ultrahigh-$n$ optical metasurfaces over a large area (**Fig. 4F**), promoting rapid uptake of this self-assembly strategy in the prototyping of metasurface/metamaterials for experts and non-experts alike.

## 3. Conclusions

Taken together, an integrative pipeline spanning the design of high-$n$ colloidal metasurfaces, assembly of polyhedra, and optical characterization has been proposed to achieve a record-high $n$ of 10-plus at optical frequencies. We systematically exploited that assemblies of Au polyhedra can outperform their spherical counterparts, ensuring a stronger capacitive coupling between resonantly induced EDs. With the given size of Au polyhedra, the deterministic control over $M_w$ of the PSSH organic ligand enabled selective reconfiguration of the nanogap. This, in turn, further elucidated the importance of a capacitive ED coupling-based underpinning mechanism for achieving an unnaturally high $n$. Notably, the achieved $n$ of 10.12 in our work was out of reach, as emphasized in **Fig. 4E**. This breakthrough has the potential to reshape various transformative applications, such as optoelectronic devices like solar cells, whose performance (e.g., absorption limit) depends on the $n$.

## Acknowledgments

**Funding**: This work was supported by the National Research Foundation (NRF) of Korea grant (NRF-2022M3H4A1A02074314 and NRF-RS-2023-00272363), Samsung Research Funding & Incubation Center for Future Technology grant (SRFC-MA2301-20), the KIST Institutional Program (2V09840-23-P023), and Korea University grant.

**Author contribution**: S.L. conceived the original ideas. N.K. synthesized colloids, assembled them, and measured optical properties. J.H.H. designed the experiments and theoretically analyzed refractive index. Y.D.C. analytically solved DLVO theory and analyzed colloidal interaction potentials. S.H.P. ran HOOMD. K.H.R. and J.L. contributed to the synthesis of Au NPs and microscopic analyses (SEM and TEM) of them. H.H.K. measured reflection spectra. S.L. supervised the project. The manuscript is mainly written by Y.D.C. and S.L. with the contributions of all authors.

**Competing interests**: The authors declare no conflicts of interest.

**Data and materials availability**: All data are available in the main text or the supplementary materials.

## Supplementary Materials

Materials and Methods

Supplementary Text

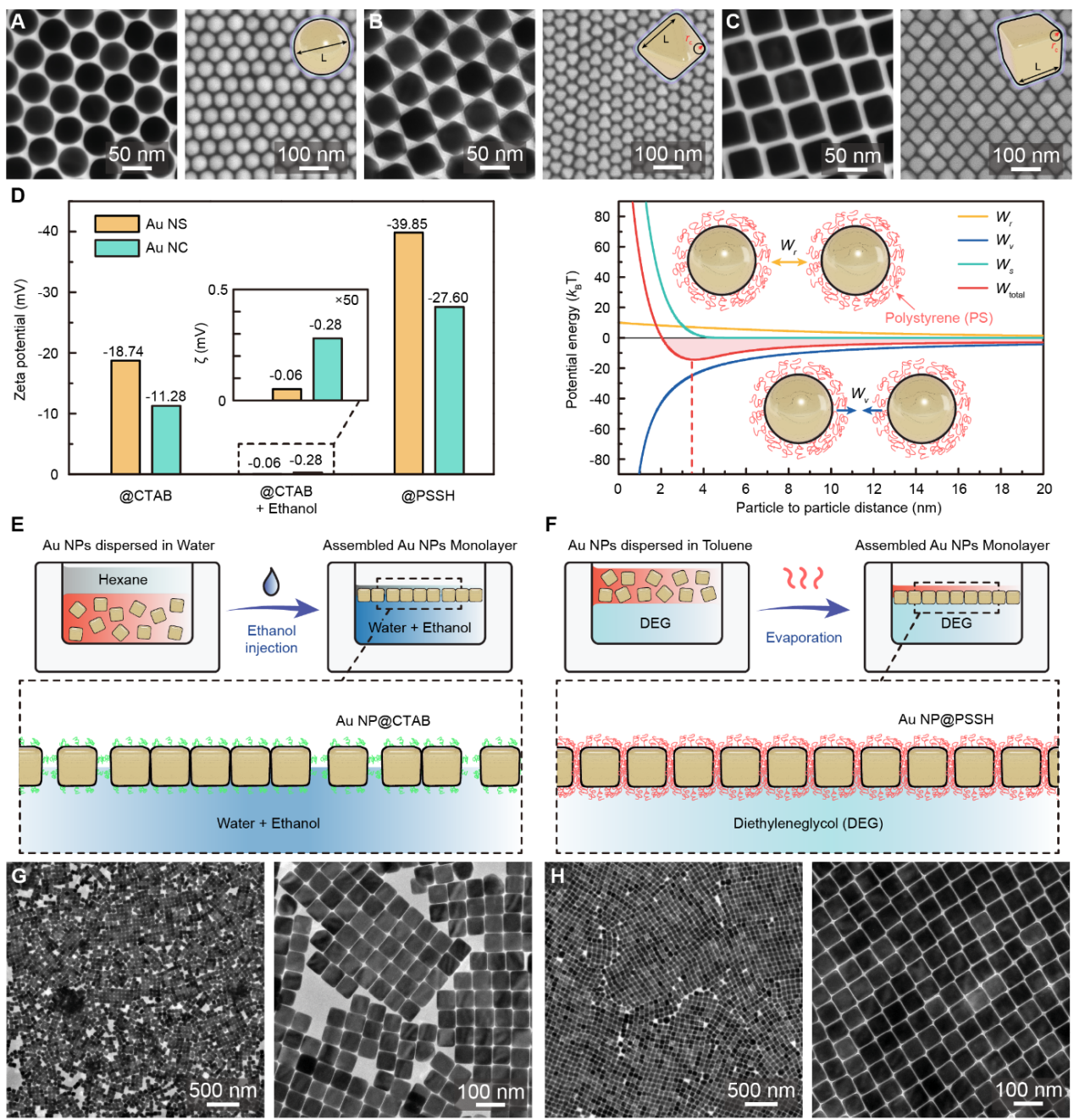

**Fig. 1. Systematic investigation of gold nanoparticles (Au NPs) synthesis, characterization, and assembly.** (A-C) electron microscopy images of uniformly synthesized Au NPs: transmission electron microscopy (TEM) images (left panel) and scanning electron microscope (SEM) images (right panel) with their characteristic lengths (L) and radius of curvature ($r_c$) for (A) Au nanospheres (Au NS), L=53.25±1.55 nm, (B) Au nanooctahedra (Au NO), L=52.95±2.21 nm, $r_c$≈8.5 nm and (C) Au nanocubes (Au NC), L=53.10±2.35 nm, $r_c$ ≈ 15 nm. (**D**) left panel, Zeta potential measurements for various ligands and environmental conditions on Au NPs: CTAB-coated without ethanol injection, CTAB-coated with ethanol injection, and PSSH-coated Au NS (yellow bar) and Au NC (green bar), and (right panel) a calculated Derjaguin-Landau-Verwey-Overbeek (DLVO) potential plot for PSSH-coated Au NPs. (**E-F**) schematic illustrations of two distinct interfacial assembly methods for 2D confined assembly of Au NPs (E) at the water-hexane (oil) interface and (F) at the diethylene glycol (DEG)-toluene interface. (**G-H**) Featured TEM images of assembled Au NCs, demonstrating the impact of different interfacial assembly methods for (G) water-oil interfacial assembly and (H) DEG-toluene interfacial assembly with low-magnification TEM images (left panel) and high-magnification TEM images (right panel), respectively.

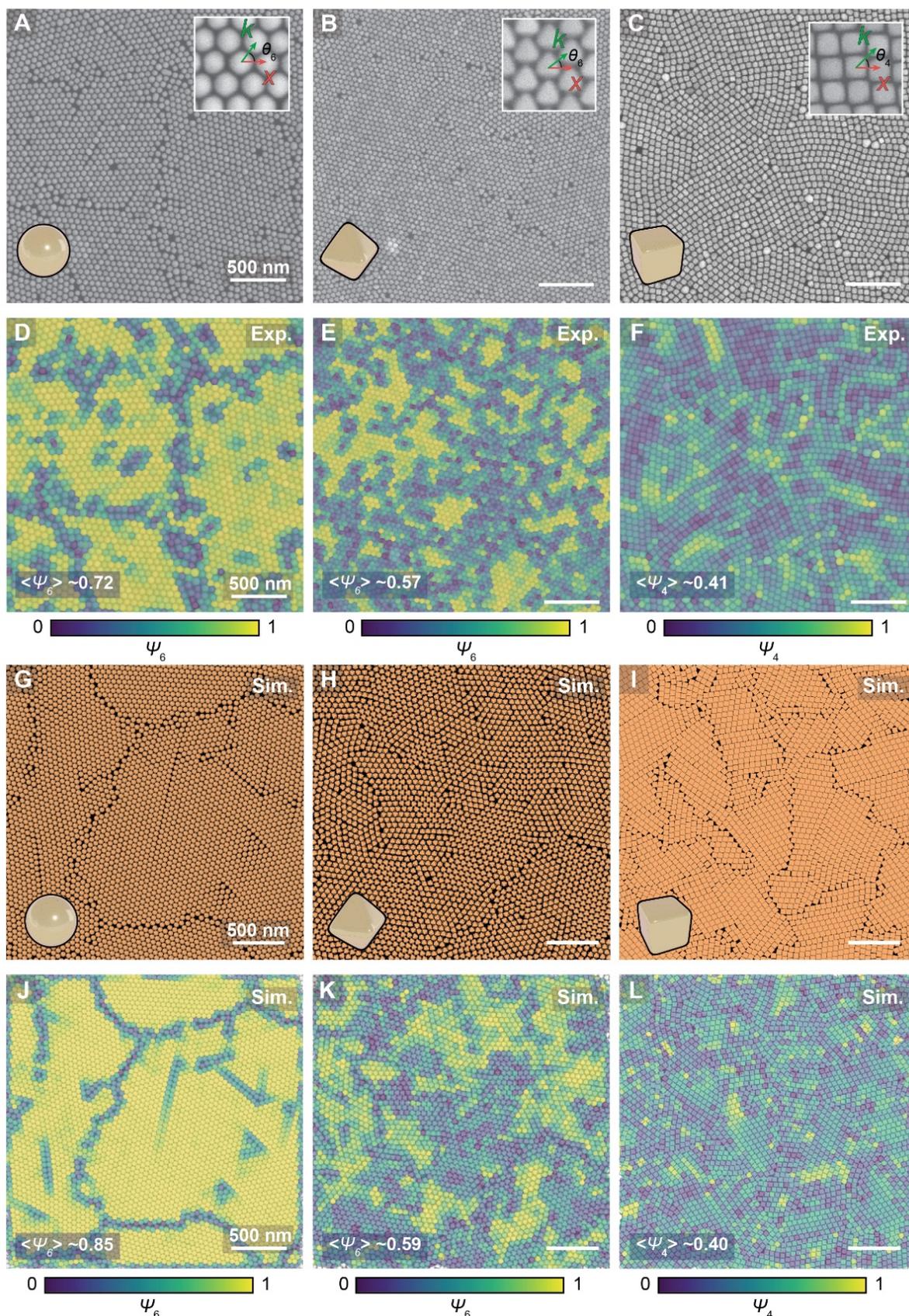

**Fig. 2. Crystallinity evaluation of assembled Au NPs using Voronoi diagrams.** (**A-C**) SEM images of assembled Au NPs for (A) Au NS, (B) Au NO, and (C) Au NC. (**D-F**) Voronoi diagrams measuring $\Psi_n$ based on the SEM images where $n$ is the $n$-fold symmetry according to the 2D lattice type for (D) Au NS, $\langle\Psi_6\rangle$=0.72, (E) Au NO, $\langle\Psi_6\rangle$=0.57, and (F) Au NC, $\langle\Psi_4\rangle$=0.41. (**G-I**) HOOMD simulation results for (G) Au NS, (H) Au NO, and (I) Au NC. (**J-L**) Voronoi diagrams measuring $\Psi_n$ based on the HOOMD images for (J) Au NS, $\langle\Psi_6\rangle$=0.85, (K) Au NO, $\langle\Psi_6\rangle$=0.59, and (L) Au NC, $\langle\Psi_4\rangle$=0.40.

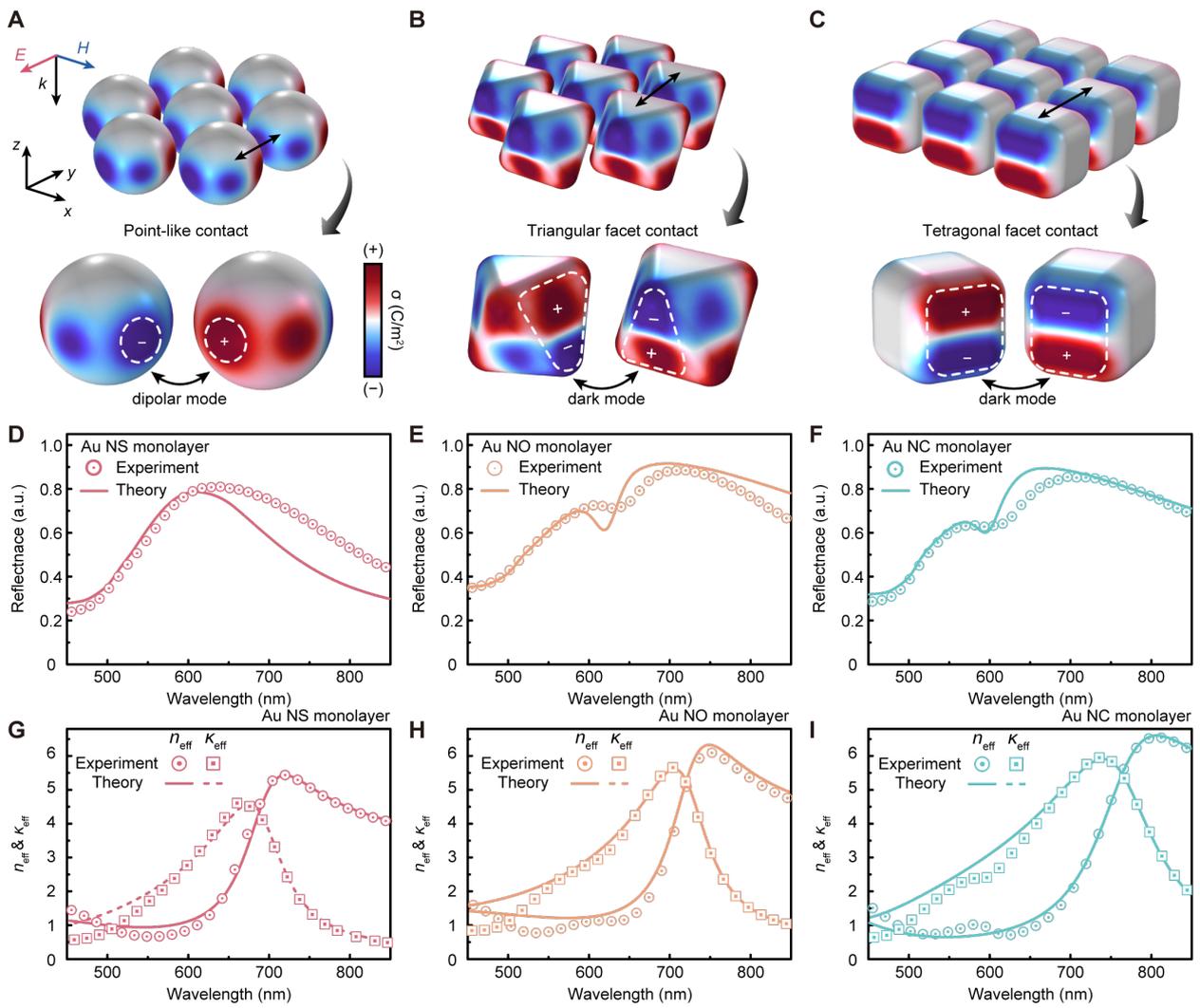

**Fig. 3. Comprehensive analysis of resonant behaviors and reflective attributes in assembled Au NP monolayers.**
(**A-C**) The illustrations of simulation results demonstrate the local surface charge distributions on closely packed Au NPs at resonance frequencies, which elucidate the Lorentzian resonant oscillatory behaviors and dark-mode resonances for (A) Au NS, (B) Au NO, and (C) Au NC monolayers. Au NS monolayer shows the bright mode, while Au NO and Au NC monolayers depict the dark-mode resonances, which are characterized by the notable dips in their respective spectra. (**D-F**) The bright field normal reflection spectra for the Au NP monolayers for (D) Au NS, (E) Au NO, and (F) Au NC, with measured spectra (dotted lines) and numerical predictions spectra (solid lines). These spectra validate the consistency of the nanogaps at approximately 6 nm for each monolayer. (**G-I**) The comparative analyses of refractive index spectra ($n$-dispersions) for the Au NP monolayers methodically correlate the refractive index dispersions with the theoretical Lorentz oscillator model for (G) Au NS, (H) Au NO, and (I) Au NC monolayers, as measured by ellipsometry spectrometer (dotted lines) and theoretical fitting curve (solid lines).

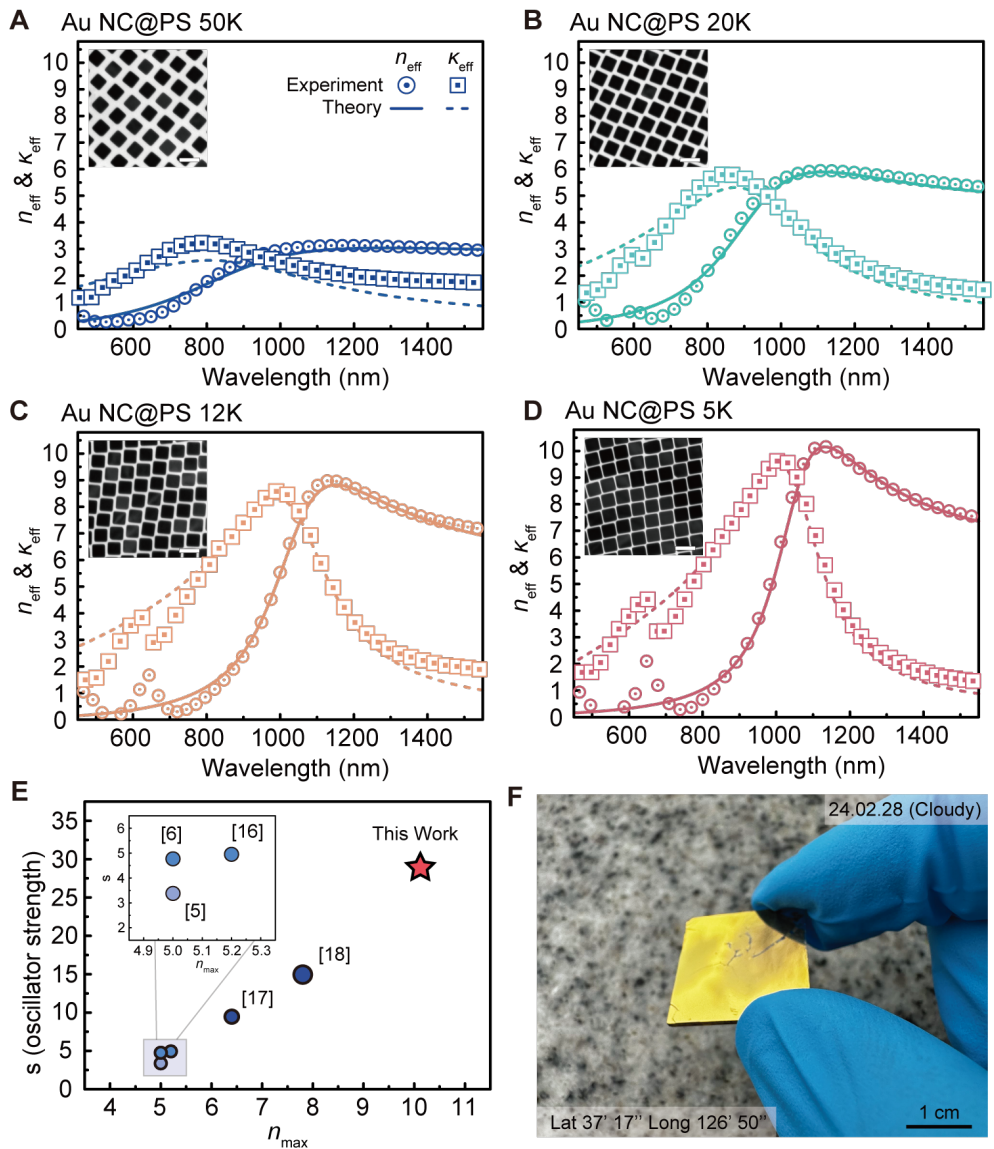

**Fig. 4. Molecular weight dependence of the refractive index of Au NC monolayers for enhanced maximum refractive index ($n_{max}$) and oscillator strength ($s$).** (**A-D**) Illustration of how varying the molecular weight ($M_w$) of PSSH affects the size of the nanogap and the capacitive coupling in 60 nm Au NC monolayers with $M_w$ of PSSH for (A) 50K, (B) 40K, (C) 12K, and (D) 5K. Increasing the $M_w$ of PSSH from 5K to 50K while maintaining the size of the Au NCs constant at 60 nm leads to an increase in the nanogap from 5 nm to 10 nm, with the interfacial area between Au NCs remaining nearly unchanged. (**E**) Comparison of $s$ and $n_{max}$ values measured in this work (star mark) with measurements from other self-assembled nanoparticle metamaterials (blue circles) (**F**) The photograph depicts the crack-free assembly of the monolayer, facilitating high reflectivity.

# Achieving Optical Refractive Index of 10-Plus by Colloidal Self-Assembly

*NaYeoun Kim*[1,+], *Ji-Hyeok Huh*[1,2,+], *YongDeok Cho*[1,+], *Sung Hun Park*[1], *Hyeon Ho Kim*[1], *Kyung Hun Rho*[1], *Jaewon Lee*[1], and *Seungwoo Lee*[1,3,4*]

[1]KU-KIST Graduate School of Converging Science and Technology, Korea University, Seoul 02841, Republic of Korea

[2]Department of Applied Physics, Hanyang University, Ansan 15588, Republic of Korea

[3]Department of Integrated Energy Engineering (College of Engineering), Department of Biomicrosystem Technology, and KU Photonics Center, Korea University, Seoul 02841, Republic of Korea

[4]Center for Opto-Electronic Materials and Devices, Post-Silicon Semiconductor Institute, Korea Institute of Science and Technology (KIST), Seoul 02792, Republic of Korea

*Email: seungwoo@korea.ac.kr

Keywords: Colloids, Polyhedra, Self-assembly, Mata-atoms, Electric polarization

[+]Equally contributed to this work

## Contents:

1. **Synthesis and characterizations of gold (Au) colloids**
2. **Measurements of Zeta potentials**
3. **DLVO theory analysis of Au nanoparticle self-assembly**
4. **Quantitation of the size distribution of Au NPs from the measured TEM images**
5. **UV-VIS absorption spectrometer for proving ligand exchange**
6. **Assembly of Au colloids with shorter PSSH (**thiol-terminated polystyrene**) (Mw of 2K)**
7. **Coordination number in a 2D assembled monolayer.**
8. **Monte Carlo (HPMC) simulations on colloidal assembly**
9. **Normal reflection spectroscopic analyses**
10. **Ellipsometry measurements**
11. **Analysis of particle assemblies using the Voronoi diagram**
12. **Numerical electromagnetic simulations**
13. **Analytic determination of the complex dielectric functions (n-dispersions)**

# 1. Synthesis and characterizations of gold (Au) colloids

*Au nanorod (Au NR) synthesis*

Seed-mediated growth of Au seeds was synthesized from the iterative reductive growth and oxidative dissolution procedure. First, to synthesize the seed, gold (III) chloride trihydrate (125 µL, 10 mM) was injected into aqueous hexadecyltrimethylammonium bromide (CTAB) solution (5 mL, 100 mM). Then, add ice-cold sodium borohydride solution (300 µL, 10 mM) to the stirring solution rapidly. To initiate seed nucleation, Au seed solution was left untouched in a 28 °C water bath for 30 min. Then, Au NRs were grown by adding gold (III) chloride trihydrate (10 mL, 10 mM), silver nitrate (1.8 mL, 10 mM), L-ascorbic acid (1.14 mL, 100 mM), and Au seeds (240 µL) into a CTAB aqueous solution (200 mL, 100 mM). Synthesized Au NR solution was left untouched in a 28 °C water bath for 2 hours. After 2 hours, it was washed with 50 mM CTAB aqueous solution. The as-synthesized Au NRs were redispersed in a CTAB aqueous solution (50 mM) and adjusted to 2 optical density (OD) concentrations. Au NRs were injected into a final $HAuCl_4$ solution with a concentration of 60 µM and gently stirred for 4 hours at 40 °C. During stirring, the tip of the Au NRs was selectively etched, and then finally we can synthesize the single crystal eAu NR (etched Au NRs). The eAu NR solutions were centrifuged with CPC (100 mM) to terminate the reaction.

*Au nanosphere (Au NS) synthesis*

To synthesize concave rhombic dodecahedron (CRD) Au NPs, Au (III) chloride trihydrate (350 µL, 10 mM), L-ascorbic acid (4.5 mL, 100 mM), and eAu NRs (6 mL, 1 OD concentration) was sequentially injected to CPC solution (20 mL, 20 mM) and stirred for 15 min sequentially. The CRD solutions were washed and redispersed in CTAB solution (50 mM) and adjusted 1 OD concentration. The CRD solution was brought to a final Au (III) chloride trihydrate concentration of 60 µM and after 2 hours, etched CRD (eCRD) was synthesized and redispersed in CPC solution (100 mM). Through iterative reduction and oxidation reactions, uniform Au Nanospheres (NSs) were synthesized. The size of Au NSs was deterministically tuned by controlling the concentration of eCRD.

*Au nano-octahedra (Au NO) synthesis*

To synthesize nanooctahedron (NOs), Au (III) chloride trihydrate (100 µL, 10 mM), ascorbic acid (AA 13 µL, 100 mM), and eCRD (1 OD, 300 µL) was injected into CPC aqueous solution (5 mL, 100 mM). Then, this solution was stirred for 30 min and centrifuged at 2100 rcf, 10 min.

*Au nanocube (Au NC) synthesis*

To synthesis nanocubes (NCs), KBr (500 µL, 100 mM), Au (III) chloride trihydrate (100 µL, 10 mM), ascorbic acid (AA, 150 µL, 100 mM), and eCRD (1 OD, 300 µL) were sequentially injected into aqueous CPC solution (5 mL, 100 mM). Then, this solution was stirred for 1 hour and centrifuged at 1700 rcf for 10 min.

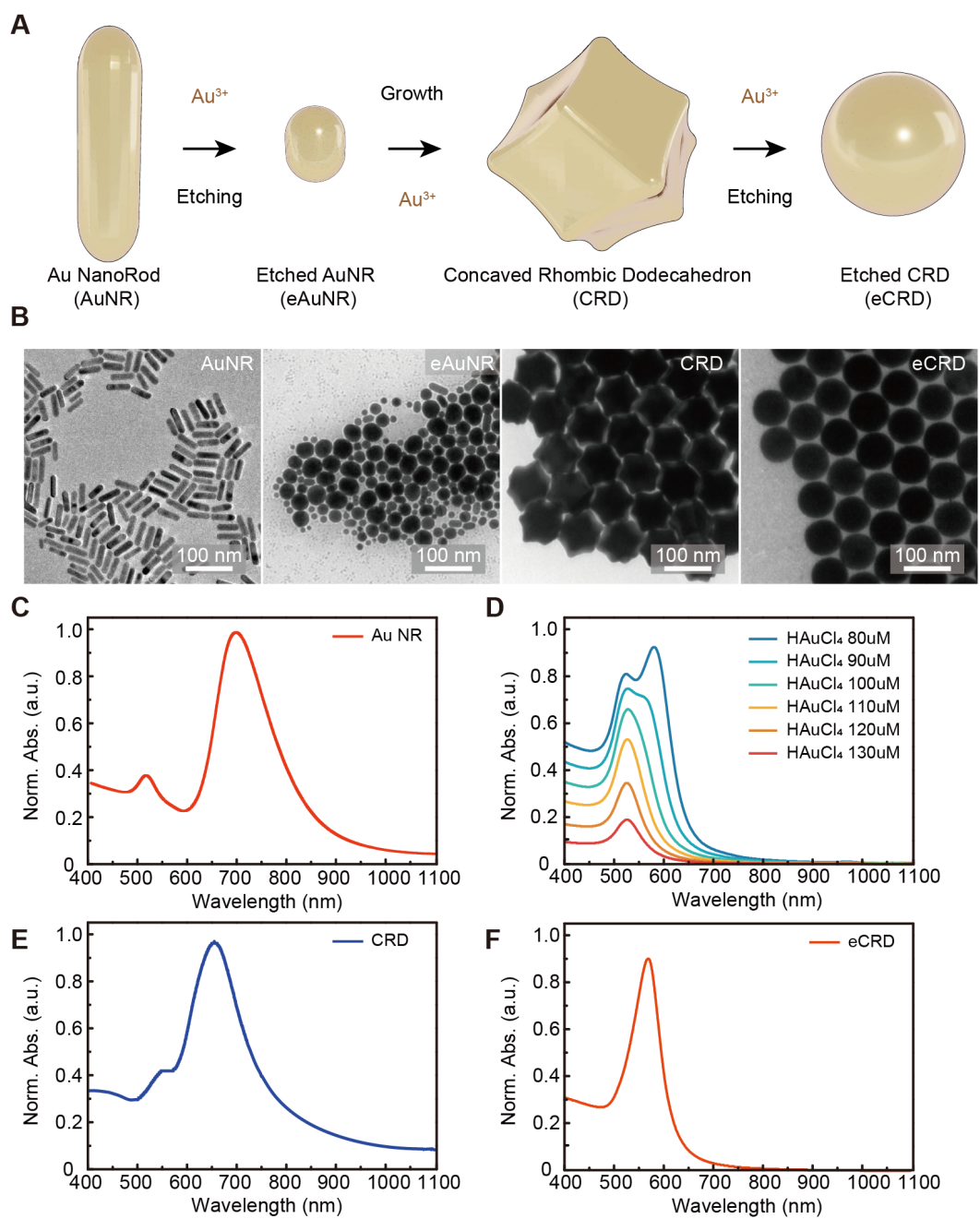

**Figure S1**. (**A**) Schematic illustration of the synthesis of uniform gold nanoparticles. The diagram outlines the synthetic protocol leading to highly uniform etched gold nanoparticles. Initially, synthesized gold nanorods (Au NRs) undergo oxidative dissolution in the presence of gold (III) chloride ions. These eAu NRs are then regrown into concave rhombic dodecahedra (CRD) by adding gold (III) chloride. To synthesize the final product of etched gold nanoparticles, the CRD nanoparticles undergo a final etching process. This oxidative dissolution is iteratively performed to precisely control the final size of the gold nanoparticles. (**B**) TEM images of AuNPs that matched to the schematic illustration (left to right panel, Au NRs, eAu NR, CRD, and eCRD) (**C**) UV/Vis spectrum of as-synthesized Au NRs. (**D**) UV/Vis spectrum of eAu NRs relative to the concentration of $Au^{3+}$. (**E**) UV/Vis spectrum of CRD. (**F**) UV/Vis spectrum of eCRD.

## 2. Measurements of Zeta potentials

The Zeta potential of Au NPs was analyzed using a Zeta-potential Analyzer (OTSUKA, ELSZ2000).

## 3. DLVO theory analysis of Au nanoparticle self-assembly

To rationalize the assembly of Au NPs into the close-packed monolayers, we used the Derjaguin-Landau-Verwey-Overbeek (DLVO) theory to numerically calculate the intermolecular forces between Au NPs. The total net force acting on the nearest two colloids can be expressed with the summation of the attractive van der Waals ($V_{vdw}$) force and repulsive electrostatic double-layer with polymeric steric forces as eq. (1).

$$V_{DLVO} = V_{vdw} + V_{coul} + V_{poly} \quad \text{(eq. 1)}$$

Our model has been set to have $R_{Au}$ = 50 nm with 5K PSSH polymeric brush grafted on the surface of Au NPs. The steric hindrance for polymeric repulsion potential was estimated using Alexander-de Gennes' polymeric brush model(*1*) and the Milner, Witten, and Cates (MWC) model(*2*)

$$\frac{V_p}{k_b T} = \frac{16\pi r L^2 \sigma^{3/2}}{35} \left[ 28\left(\left(\frac{2L}{h}\right)^{\frac{1}{4}} - 1\right) + \frac{20}{11}\left(1 - \left(\frac{h}{2L}\right)^{\frac{11}{4}}\right) + 12\left(\frac{h}{2L} - 1\right)\right] \quad (0 \leq h \leq 2L) \quad \text{(eq.2)}$$

where the $r$ is the radius of gold colloids, the $L$ is the thickness of the 5K PSSH polymeric brush (estimated to be around 3 nm), $\sigma$ is the surface density of PSSH grafted on the Au NP surface (measured experimentally via dissolution of ligand as 0.2 per /nm$^2$).

Along with polymeric brush-driven repulsion forces, the electrostatic repulsion force also plays a key element in colloidal behavior. With the help of the Poisson-Boltzmann distribution, the electrostatic repulsive force has been estimated with the following equation.

$$\frac{V_{EDL}}{k_b T} = 2\pi r * \psi_0^2 e^{-\frac{r}{d}} * \varepsilon_r \varepsilon_0 \quad \text{(eq. 3)}$$

where $\varepsilon_0$, $\varepsilon_r$, $r$, $\psi_0$, and $\lambda_d$ correspond to the permittivity of solvent, the relative permittivity of toluene, radius of colloids, surface potential of colloids, and Debye screening length, respectively. Herein, 2.38 of $\varepsilon_r$, 20 nm of $r$, 40.5 mV of $\psi_0$, and 9.6 nm of $\lambda_d$ were used, which were obtained from experimental measurements.

On the other hand, the attractive vdW force can be expressed as follows eq. 4

$$V_{vdW} = -\frac{A_H * r}{12H} \quad \text{(eq. 4)}$$

where, $A_H$ is the Hamaker constant of Au NPs dispersed in toluene solvent, which was estimated as 160 x 10$^{-21}$ J (*3*)

## 4. Quantitation of the size distribution of Au NPs from the measured TEM images

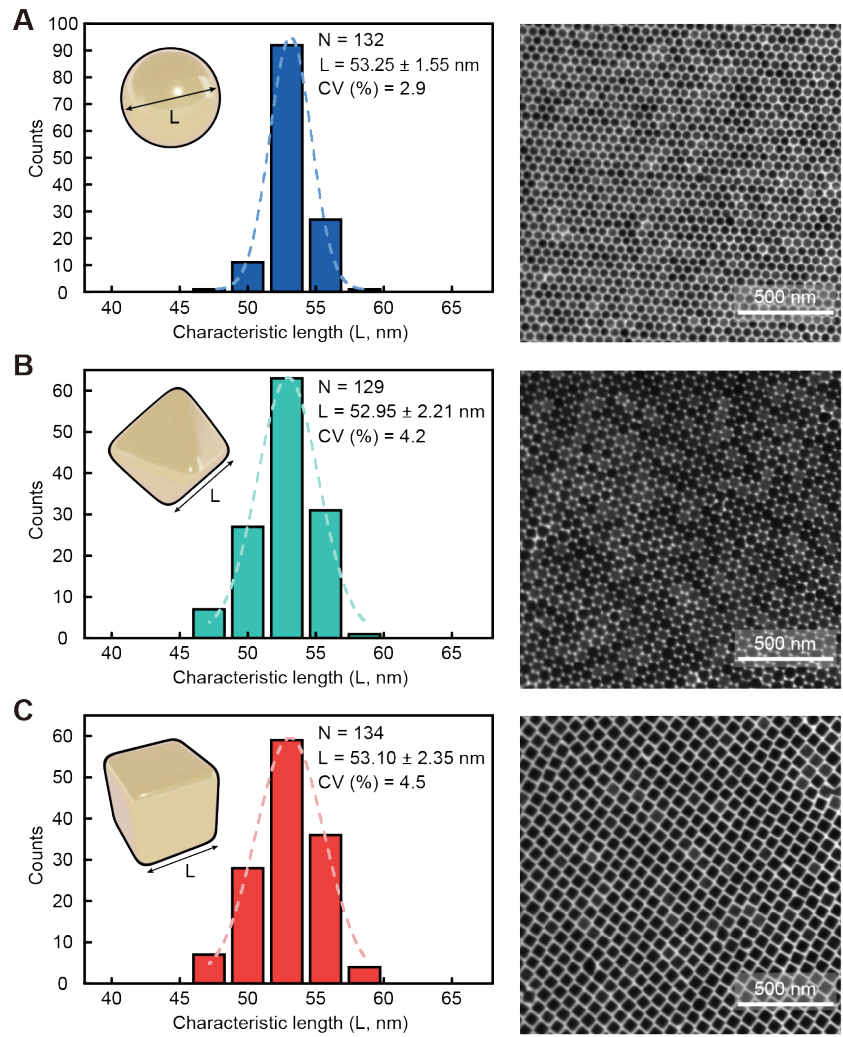

**Figure S2.** The statistical size distribution analysis of gold nanoparticle is presented for (**A**) Au NSs, (**B**) Au NO, (**C**) Au NCs monolayers respectively. To accurately quantify the size distribution of AuNPs, we employed Matlab's Image Analysis Toolbox to analyze TEM images displayed on the right side of the panel. Each row presents size distribution of Au NSs, Au NO, and Au NCs based on approximately 130 samples of particles.

## 5. UV-VIS absorption spectrometer for proving ligand exchange

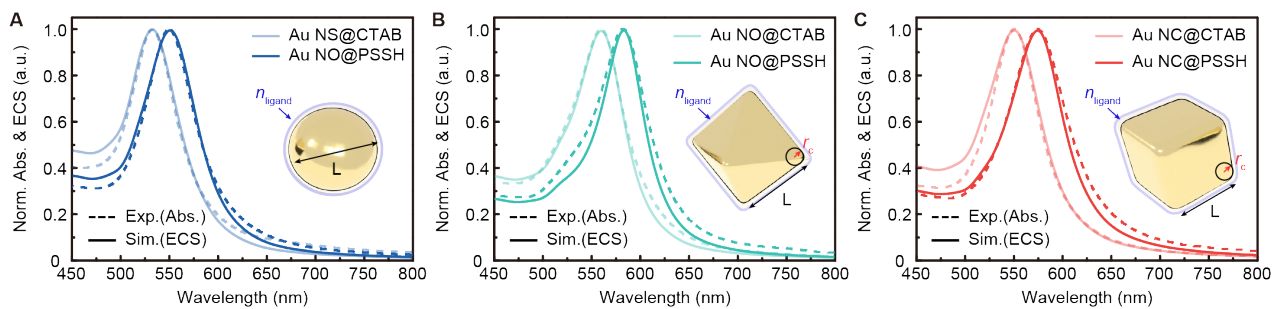

**Figure S3.** Extinction spectrum analysis of gold nanoparticles with surface coatings. The panels (**A**) Au NS, (**B**) Au NO, and (**C**) Au NC show the extinction spectra for gold nanoparticles coated with CTAB (represented by light lines) and PSSH (depicted by bold lines), elucidating the impact of surface coatings on their optical properties. The spectra comprise both numerical simulations (solid lines) and experimental observations (dotted lines). The observed spectral shift across the nanoparticles can be attributed to the differential refractive indices of the coatings, with PSSH demonstrating a higher refractive index compared to CTAB.

## 6. Assembly of Au colloids with shorter PSSH (Mw of 2K)

Through the ligand exchange process, we implemented a PSSH polymer with a molecular weight of 2,000 g/mol (Mn) and 2,300 g/mol (Mw). This shorter PSSH polymer can reduce the interparticle distance; consequently, further enhancing capacitive coupling between Au NPs. However, this use of relatively shorter PSSH polymer as the ligand inevitably decreased the steric hindrance between the Au NPs, resulting in insufficient force to overcome the van der Waals forces. Experimental results demonstrated that during the THF evaporation process, 50 nm Au NCs with 2K PSSH led to the irreversible aggregation.

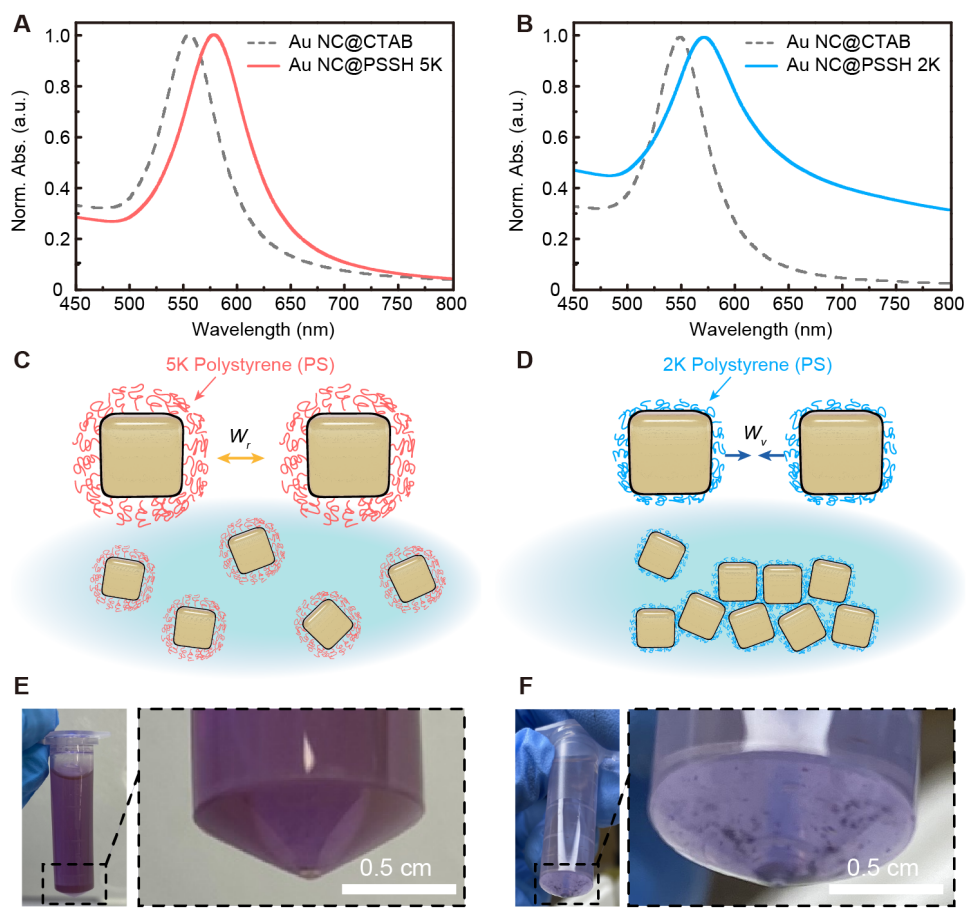

**Figure S4**. UV/Vis spectral analysis and schematic representations of PSSH-functionalized AuNCs. (**A**) UV/Vis spectrum comparing AuNC@PSSH 5K with AuNC@CTAB. The higher refractive index of the PSSH polymer induces a red shift in the resonance peak of AuNC, suggesting changes in the optical properties due to the polymer coating. (**B**) UV/Vis spectrum of AuNC@PSSH 2K and AuNC@CTAB. Spectrum broadening is observed, indicating the partial aggregation of AuNCs attributed to the reduced length of the PSSH polymer, which affects the stability and dispersion of nanoparticles. (**C-D**) Schematic illustrations of PSSH-functionalized AuNCs: (**C**) with 5K PSSH and (**D**) with 2K PSSH, respectively. These diagrams depict the molecular configuration and relative polymer lengths surrounding the AuNCs. (**E-F**) Photographs of 2 mL tubes containing PSSH-functionalized AuNCs: (**E**) with 5K PSSH and (**F**) with 2K PSSH. Due to the reduced Debye length in the 2K PSSH variant, AuNC@PSSH 2K exhibits visible aggregation, highlighting the impact of polymer chain length on nanoparticle assembly and stability.

## 7. Coordination number in a 2D assembled monolayer

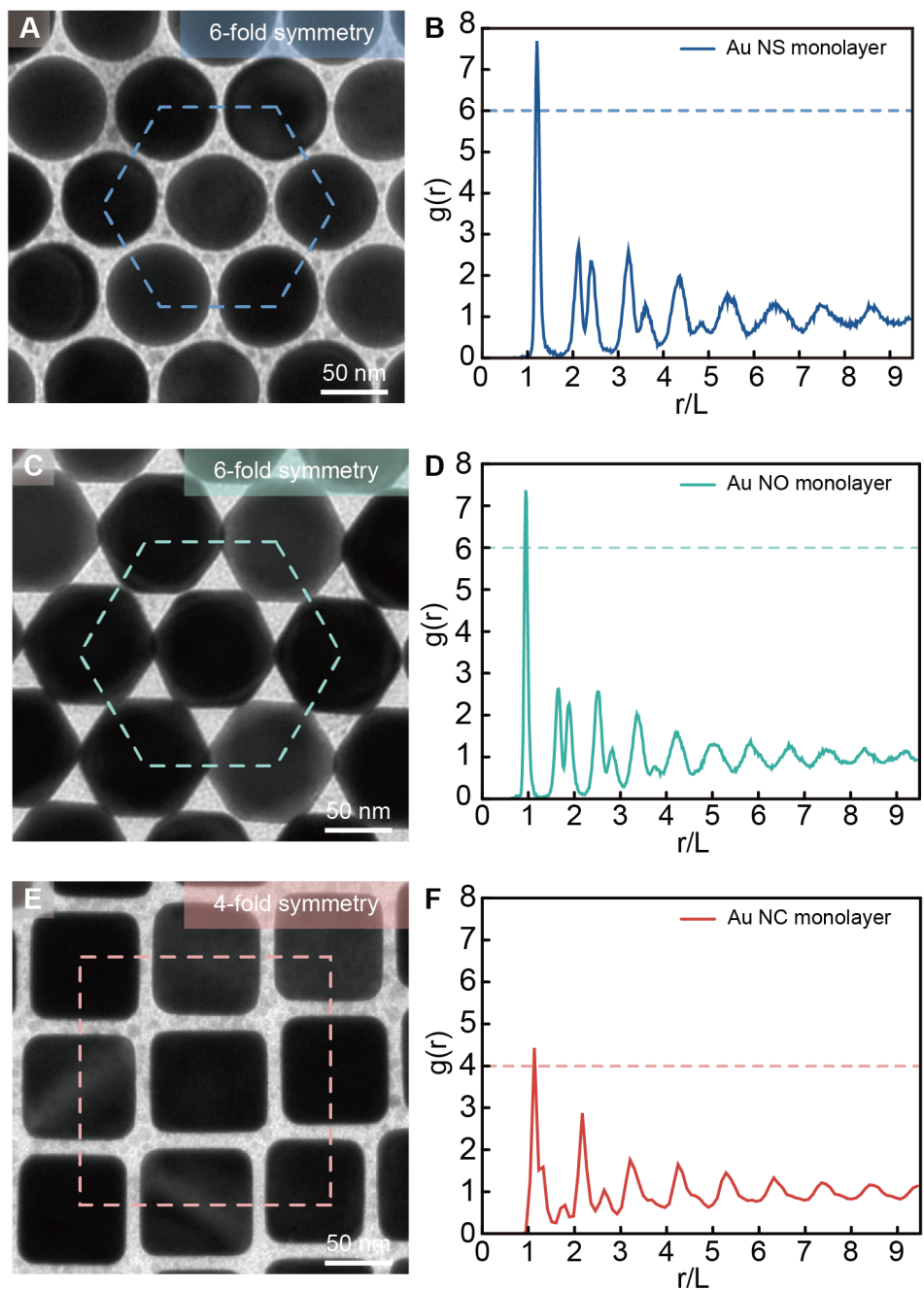

**Figure S5** The analysis of the coordination number in a self-assembled monolayer of gold nanoparticles. (**A**) TEM image showing a AuNS monolayer with a hexagonal close-packed arrangement, indicating that each nanoparticle is surrounded by six nearest neighbor Au nanoparticle (i.e., 6-fold symmetry). (**B**) Radial distribution function (RDF) for Au monolayer analyzed using aforementioned TEM image. The RDF exhibits distinctive peaks at a coordination number of 6, correlating with the 6-fold symmetry observed in the left panel of TEM image. (**C**) TEM image of AuNO monolayer. (**D**) RDF analysis of following TEM image of AuNO monolayer, indicating 6-fold symmetry. (**E**) TEM image of AuNC monolayer. (**F**) RDF analysis of following TEM image of AuNC with distinctive peak, indicating 4-fold symmetry of self-assembled monolayer.

## 8. Monte Carlo (HPMC) simulations on colloidal assembly

In this study, all numerical simulations were performed using the free software HOOMD-blue v2.9.7 (stable version)(*4, 5*). We performed each simulation with N=8,000 particles of AuNSs, AuNO, and AuNCs in a hexahedral box together with periodic boundary conditions. Initially, all identical particles were arranged in a simple cubic lattice. The first objective was to demonstrate the experimental system of 2D confinement at the water-oil interface. To achieve this, the length of the simulation box along the z-axis was rapidly compressed until the spatial distribution of all particles was confined to the xy-plane. Subsequently, potential walls that prevent the particles from entering specific areas were placed at the top and bottom of the simulation box.

## 9. Normal reflection spectroscopic analyses

All self-assembled AuNSs, AuNO, and AuNCs monolayers were analyzed using a custom-built optical instrument. Specifically, the normal reflection spectra were collected using an optical microscope (Nikon ECLIPSE LV100ND) equipped with an image sensor (Nikon DS-Fi3) and an imaging spectrometer (IsoPlane, Princeton Instruments) coupled with a CCD camera (PIXIS-400B). A broadband visible light from a 12 V/100 mW halogen lamp was directed onto the monolayer through a 5x objective lens with a numerical aperture (NA) of 0.3. The light source used in the normal reflection analyses was polarized along the s-pol. The normally reflected light from Au NP monolayers was collected by using a 50x objective lens with an NA of 0.8 and spectrally analyzed.

## 10. Ellipsometry measurements

The effective refractive index of Au NP monolayers was measured using a commercially available ellipsometry spectrometer (Elli-SEUN-AM12, Ellipso-technology, Korea) over a range of wavelengths from UV and visible to near IR (220~1650 nm). The measurements were taken using the reflective mode, and the samples were illuminated with light at a 70.35° incidence angle and a beam spot size of 1.5 mm × 4.0 mm. The covered area of each Au NP monolayer was prepared more than 1 cm² due to the size of the beam spot.

## 11. Analysis of particle assemblies using the Voronoi diagram

In this study, we quantitatively analyzed the crystalline integrity of Au NP monolayers by using Voronoi diagram. For Au NS and Au NO, we used a hexagonal lattice, while for Au NC, we utilized a square lattice, reflecting the inherent geometric distinctions in particle morphology and their influence on the local and global assembly patterns. The bond-orientational order parameter, $\Psi_n$, was employed to quantify the degree of ordering within each Au NP monolayer, where $n$ is the coordination number: 6 for hexagonal and 4 for tetragonal lattices. This parameter is defined as:

$$\Psi_n = \frac{1}{N_n} \sum_k \exp(in\theta_k)$$

Here, $N_n$ represents the number of nearest neighbors, $k$ is the particle of interest identified through a Voronoi diagram, and $\theta_k$ is the angle between the positive x-axis and the **k**-vector connecting the particle to each of its neighbors (see inset of **Fig. 2A−C**). The $\Psi_n$ value varies from 0, indicating order, to 1, denoting perfect lattice order. Additionally, the local orientation, $\theta_n$, was calculated from the argument of $\Psi_n$, providing insights into the directional properties of the assemblies:

$$\theta_n = \arg(\Psi_n)$$

For hexagonal systems, $\theta_6$ indicates the local orientation with a sixfold symmetry, limited to 60°, whereas, for tetragonal systems, $\theta_4$ indicates the local orientation within a fourfold symmetry, limited to 90°. The Voronoi diagram technique was integral in plotting these parameters, visually representing particle organization and grain boundary identification. Regions with low $\Psi_n$ values adjacent to ordered grains exhibiting distinct $\theta_n$ values were identified as grain boundaries, illustrating the capacity of this algorithm in discerning the microstructural features of the nanoparticle assemblies. This strategy emphasizes the significance of lattice symmetry and particle shape in analyzing the assembly of NP monolayers, providing a comprehensive view of their lattice integrity and assembly behavior.

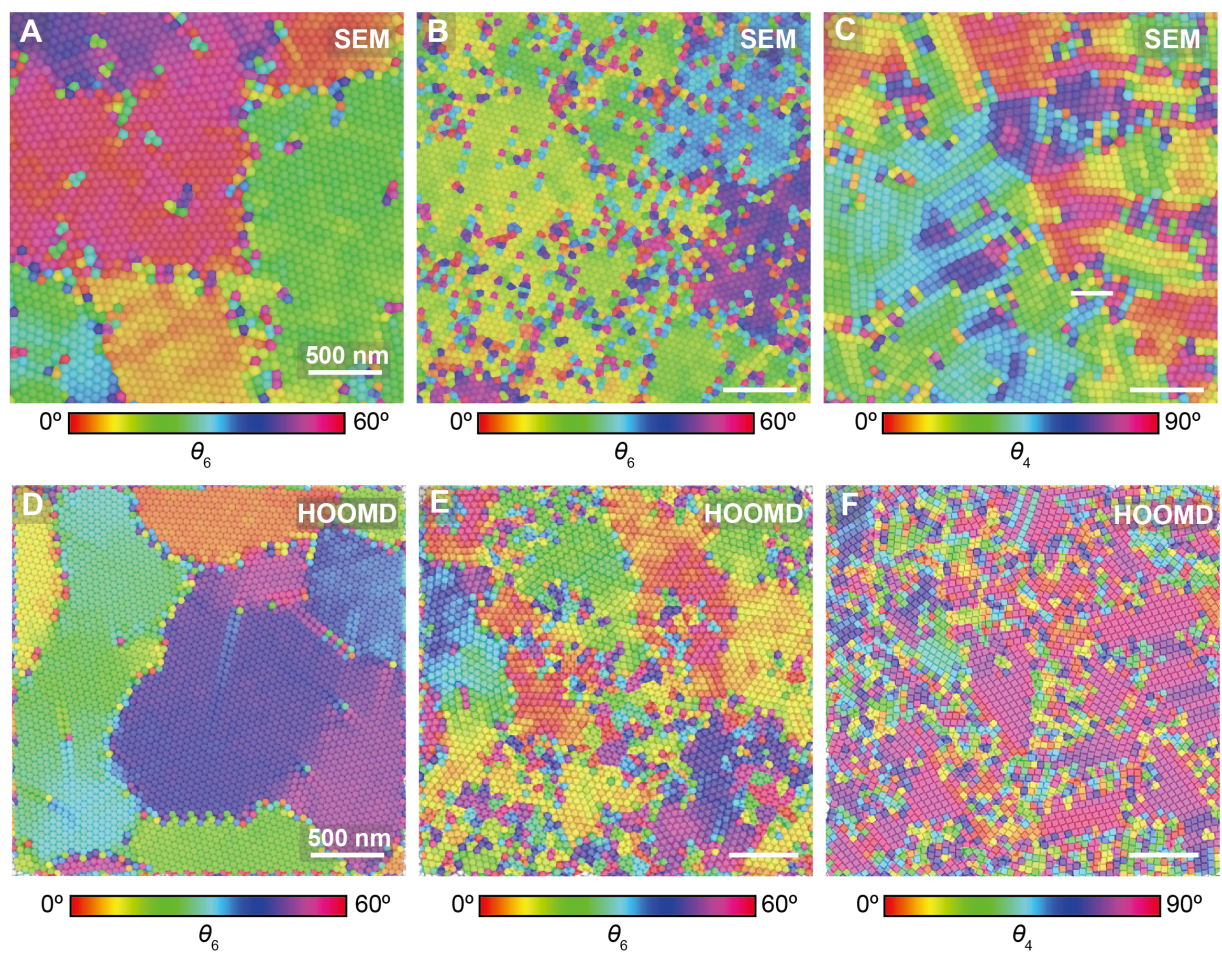

**Figure S6** Visualization of local orientation angles (θ) in Au NP assemblies. **(A-C)** the distribution of θ values for Au nanoparticle shapes: AuNS, AuNO, and AuNC derived from SEM images from **Fig. 2A-C**, illustrating the experimental local orientation angles within the assemblies. **(D-F)** the θ distributions for the same Au NP shapes, but these are obtained from HOOMD simulation data. The color gradient across all panels represents the orientation angle range from 0° to 60°, highlighting the directional properties of the nanoparticle assemblies as determined both experimentally and computationally. This comparative visualization demonstrates the influence of particle geometry on the assembly characteristics and validates the simulation approach corresponds to experimental findings.

## 12. Numerical electromagnetic simulations

The frequency domain 3D finite-element simulations of Au NS, Au NO, and Au NC were carried out using the commercial software package COMSOL Multiphysics (COMSOL, Inc). Initially, we verified the optical properties of an individual NP rather than the assembled monolayer. To determine the characteristic length ($L$) and radius of curvature ($r_c$) of NPs, as well as the effective thickness ($t$) and refractive index ($n_{\text{ligand}}$) of the surrounding organic ligand shell, we performed a numerical calculation. This process involved fitting the experimental UV/Vis absorbance spectrum (dotted line in **fig. S3**) with respect to the extinction cross-section of each NP (dotted line in **fig. S3**), as shown in **fig. S3**. After defining the size and shape parameters of the NPs, we proceeded to compute the reflectance spectrum and analyze the surface charge distributions of Au NP monolayers.

## 13. Analytic determination of the complex dielectric functions (*n*-dispersions)

To determine the effective refractive indices (*n*-dispersions) of the closely packed Au NP monolayers, we utilized a modified multiple Lorentz model to describe their complex dielectric functions, as follows:

$$\varepsilon(\omega) = \varepsilon_\infty + \sum_i \frac{s_i}{\omega_{0,i}^2 - \omega^2 - i\omega\gamma_{0,i}} \qquad \text{(eq. 5)}$$

In this formula, we empirically determined several parameters such as $\varepsilon_\infty$, $s_i$, $\omega_p$, $\omega_0$ and $\gamma_0$. Here, $\varepsilon_\infty$ represents the permittivity at infinite frequency, offers a baseline for the dielectric behavior in the absence of materials resonance. The index $i$ denotes the number of pole pairs, and the term $s_i = f_i\omega_{0,i}^2$, indicates the strength of $i$-th pole at resonance, where $f_i = N_i/N_{tot}$ describe the fraction of electrons participating in transitions, with $N_i$ is the number of transition electrons. $N_{tot}$ is the total number of electrons, while $\gamma_0$ indicates the reciprocal of the pole relaxation time. $\omega_p$ is plasma frequency.

In our study, a mono-polar ($i = 1$) model was employed for all samples, as illustrated in **Fig. 3A-C** and **Fig. 4A-D** due to the singlet resonance behavior in all Au NP monolayers (we neglect the shoulder peak caused by dark mode). Detailed results and data for each Au NP monolayer are listed in **Table 1** and **2**. The parameter $s_i$ corresponds to the strength of the harmonic oscillators (poles), allowing us to estimate the materials' polarization density, while $\omega_0$ and $\gamma_0$ provide the resonance frequency and the averaged damping rate for the $i$-th poles, respectively. Herein, the speed of light is assumed to be unity in our calculations.

|        | $s$  | $\omega_p$ | $\omega_0$ | $\gamma_0$ |
|--------|------|------------|------------|------------|
| Au NS  | 4.76 | 3.92 eV    | 1.79 eV    | 0.23 eV    |
| Au NO  | 6.02 | 4.21 eV    | 1.71 eV    | 0.21 eV    |
| Au NC  | 8.74 | 4.77 eV    | 1.62 eV    | 0.24 eV    |

**Table S1.** Values of the parameters used in the mono-polar fit of eq. 5 to the data of **Fig. 3**.

|          | $s$   | $\omega_p$ | $\omega_0$ | $\gamma_0$ |
|----------|-------|------------|------------|------------|
| PSSH-5K  | 29.06 | 6.31 eV    | 1.17 eV    | 0.23 eV    |
| PSSH-12K | 27.68 | 6.18 eV    | 1.18 eV    | 0.30 eV    |
| PSSH-20K | 17.97 | 5.48 eV    | 1.29 eV    | 0.50 eV    |
| PSSH-50K | 7.38  | 3.90 eV    | 1.43 eV    | 1.00 eV    |

**Table S2.** Values of the parameters used in the mono-polar fit of eq. 5 to the data of **Fig. 4**.